# Synergistic Influence of Surface Sanding Technique and Multi-Walled Carbon Nanotube Incorporation on the Mode I Fracture Behavior of Glass Composite Laminates

## Roohollah Nazari *[1], Rashid Hakimi [2], Mohammad Daneshfar [2], Behnam Talebi [2]

[1]*TriboDynamic research laboratory, Department of Mechanics, Mathematics and Management, Politecnico of Bari, Via Orabona 5, 70125, Bari, Italy*

[2]*School of Mechanical Engineering, Iran University of Science & Technology (IUST), Narmak, Tehran 16846-13114, Iran*

**Abstract**:

This study sequentially investigates the enhancement of fracture behavior in adhesive joints. Initially, mechanical sanding techniques were explored, revealing significant improvements of 55% in fracture energy and 38% in load-bearing capacity with optimal 240-grit sanding. Subsequently, in a second stage, Multi-Walled Carbon Nanotubes (MWCNTs) were introduced at varying weight percentages (0.1%, 0.3%, and 0.5% by weight) into the adhesive joints. This combined approach aimed to synergistically examine the impact of surface pretreatments and nanoparticle integration on fracture behavior. Results highlight the potential for fracture resistance and enhanced load-carrying capabilities in reinforced specimens, with maximum load and fracture energy improvements of up to 92 % and 50 %, respectively, compared to unreinforced specimens and those subjected solely to sanding treatments. Furthermore, the fracture mode shifted from adhesive failure to cohesive substrate failure in reinforced specimens. This comprehensive investigation underscores the intricate interplay between mechanical preparation and nanomaterial incorporation in significantly improving adhesive joint performance

**Key Words:** Polymer matrix composites, nanomaterials, adhesive, fracture mechanics, multi-walled carbon nanotube

---

* **Corresponding author**
E-mail address: r.nazari@phd.poliba.it **(R. Nazari)**

## 1. Introduction

The ability to connect and integrate components is a critical aspect of the manufacturing process. Mechanical fastening, achieved through mechanical joining, is a commonly employed technique. However, there are various drawbacks associated with the use of these fasteners, including the potential for damaging drilling components, increasing the overall weight of the structure, and introducing significant localized stresses [1]. Additionally, considering the involvement of manual labor, this method is time-consuming and lacks flexibility.

Conversely, adhesive joints offer several benefits such as even distribution of stress, improved resistance to fatigue, reduced structural weight, and the capability to join dissimilar materials [2, 3]. These advantages have rendered adhesive joints highly attractive in industries like aviation, marine, and automotive. It's important to acknowledge, however, that adhesive joints are often the weakest link in a structure due to their relatively limited fracture tolerance and susceptibility to harsh environmental conditions [4].

The performance of adhesive joints can be significantly affected by various factors, including the material properties of the adhesive, the thickness of the adhesive layer, the condition of the bonding surfaces, and the design of the joint [5-7]. Ensuring adequate bonding strength, resistance to fracture, and enduring durability within adhesive joints heavily relies on the attributes of the bonding surfaces. The primary objective of surface pretreatment is to remove any surface impurities like lubricants or oils and to address vulnerable boundary layers. This process aims to enhance the bonding area by optimizing wettability, establish mechanical interlocking, and induce alterations in the surface morphology through chemical means [8]. A range of surface treatment techniques, including mechanical roughening [9,10], peel ply [11,12], chemical treatments [13], and plasma treatments [14-15], have been utilized to attain desired surface characteristics for composite lamintes.

The surface texture and roughness play a direct role in influencing how adhesive flows and spreads across the substrate surface. Optimal surface roughness can facilitate adhesive penetration into the substrate, leading to an increase in the effective bonding area [16,17]. Additionally, the concept of mechanical interlocking suggests that once the adhesive fills the gaps on the substrate surface, the cured adhesive establishes interlocking connections with the uneven surface topography of the substrate [17]. Furthermore, research indicates that the impact of mechanical interlocking is amplified with higher surface roughness, consequently resulting in enhanced bonding strength [18]. Gude et al. [19] conducted a study examining the relationship between surface attributes (specifically roughness and surface free energy) and the mechanical properties of adhesive joints. The experimental findings highlighted that mechanical interlocking primarily governed the adhesion mechanism for mode I testing. Interestingly, in the context of shear strength, the results indicated that an increase in surface roughness did not necessarily lead to an improvement, and instead, the polar component of the surface free energy exerted a more substantial influence on shear strength.

Numerous researches have investigated the impacts of diverse surface treatment approaches on adhesive-bonded joints. In the study by Kumar et al. [20], atmospheric pressure plasma treatment was employed to alter the bonding surfaces of adhesive single lap joints. The research demonstrated that this treatment method could be deemed efficient in enhancing the adhesive properties of graphite-epoxy laminate surfaces. In the work conducted by Sun et al. [21], an exploration was conducted into the impact of different patterns generated on the bonding surfaces through laser treatment on the fracture behavior of adhesively bonded joints. The outcomes demonstrated that introducing both longitudinal and transverse grooves on the bonding surfaces of the joints resulted in an elevation of the joint mode-I fracture energy value. In the investigation by Martinez-Landeros et al. [22], an examination was carried out to understand the impact of different surface pretreatment techniques, such as solvent cleaning, sanding, chemical etching, and peel ply, on the fracture characteristics of carbon fiber-reinforced composites. The findings suggested that the method of mechanical abrading yielded the most significant enhancements in terms of load-bearing capacity and fracture energy of the bonded joints.

The sanding technique, employed to prepare bonding surfaces, represents a pragmatic and straightforward approach that yields numerous advantages for enhancing the effectiveness of adhesive joints in composites. Through the regulation of surface roughness, sanding enlarges the bonding area and facilitates superior adhesive wetting. This, in turn, contributes to heightened maximum bonding strength, enhanced fracture resistance, and increased long-term durability. In a study by Yang et al. [9], the impact of sandpaper grit size and sanding direction on the tensile properties of carbon fiber-reinforced polymer (CFRP) single lap bonded joints was examined. The findings highlighted the significant influence of sandpaper grit size on surface characteristics. Moreover, it was observed that the highest strength was achieved when the surfaces were abraded in random directions.

The enhancement of bonding strength and fracture toughness in the polymer matrix of fiber-reinforced polymer composites (FRPCs) has also been explored through the incorporation of second-phase additives. These additives can vary in terms of size, material composition, and geometric characteristics. Out of the numerous options for second-phase additives, a notable focus has been on nanofillers, with a specific emphasis on carbon-based materials like carbon nanotubes (CNTs) and graphene nanoplatelets (GNPs). These nanofillers have garnered significant interest in recent times due to their ability to contribute to various toughening mechanisms during crack propagation, ultimately leading to an enhancement in the fracture resistance of polymers [23-26]. In a study by Gude et al. [27], they incorporated CNTs and carbon nanofibers (CNFs) into an epoxy adhesive. As a result, they observed significant improvements in the adhesive's fracture energy, with a 10% increase for CNTs and a 23.5% increase for CNFs. Khoramishad et al. [28] conducted a study to evaluate the improvement in the mode-I fracture characteristics of an epoxy adhesive through the introduction of MWCNTs. Their findings revealed a remarkable enhancement, showing a maximum 58.4% increase in the mode-I adhesive fracture energy upon adding 0.3 wt% MWCNTs.

While there are existing studies in the literature concerning the reinforcement of composite laminated adhesive joints [29,27], it's worth noting that the synergistic impacts resulting from the combination of surface treatment techniques and the incorporation of nanomaterials at different weight percentages on the fracture behavior of glass fiber-reinforced polymer (GFRP) joints have not been explored in previous research. Hence, the present study aims to conduct an in-depth exploration of the combined influence of the sanding method using various grit sizes (60, 240, 800) and the incorporation of CNTs at different weight percentages (0.1%, 0.3%, 0.5% wt) on the fracture behavior of GFRP joints. This research seeks to shed light on the potential synergistic effects arising from these dual factors and their impact on the mechanical performance of the adhesive joints in GFRP structures. Various surface parameters of GFRP adhesive joints were meticulously examined, encompassing factors such as surface morphology, surface roughness, surface free energy, and contact angle. The study focused on determining the fracture energy of adhesive joints. Additionally, the research delved into analyzing how surface treatment and nanoparticle addition influences the fracture behavior of these adhesive joints.

## 2. Materials and specimen manufacturing

### 2.1. Materials

The double cantilever beam (DCB) specimens were constructed using laminated substrates made from bidirectional E-glass woven fabrics. These substrates had a fiber density of 1.65 g/cm³ and a thickness of 0.33 mm. The assembly of the substrates was achieved using a two-component epoxy adhesive. The composite laminates and adhesive were composed of KER 828 epoxy resin from Kumho P&B Chemicals Inc., Korea, along with a compatible hardener. The KER 828 epoxy resin served as both the matrix for the composite laminates and the adhesive for joint fabrication. Table 1 presents the physical and chemical properties of the KER 828 epoxy resin. For this study, triethylenetetramine (TETA) was selected as the hardener for the epoxy resin. A 10% weight mixing ratio was chosen, which equates to the addition of 10 grams of TETA for every 100 grams of epoxy resin. This specific mixing ratio was carefully determined to ensure proper curing and to enhance the mechanical performance of the epoxy resin, resulting in strong adhesive properties. Additionally, the nanocomposite adhesive joints were reinforced using combinations of carbon nanofillers with varying mixing ratios. The MWCNTs employed in this research were sourced from Neutrino Co., Iran, and possessed an average diameter of 10 nm, a length of 20 µm, and a purity level of 95%.

**Table. 1.** Physical and chemical properties of KER 828 [24].

| Color | Density | Viscosity | Young's Modulus | Tensile Strength | Boiling point | Vapor Pressure |
|---|---|---|---|---|---|---|
| Pale yellow | 1160 Kg/m$^3$ | 12-14 Pa.s | 4.3 GPa | 73.5 Mpa | >200 C | <.01 Pa |

## 2.2. Fabrication of laminated composite substrates

E-glass fiber/epoxy laminated composites were employed for producing the joint substrates. The fabrication of these composite laminates involved using a hand lay-up technique with static load compression, utilizing a manufacturing mold with dimensions of $200 \times 200 \times 2.7$ mm³. The fabrics were cut to a size of $200 \times 200$ mm². To prevent the adhesion of the composite laminates to the mold surface and ensure smooth sample surfaces, mold release agent was applied to the mold surfaces. The laminate thickness was controlled by the height of the manufacturing mold, set at 2.7 mm. Each laminate consisted of eight plies of bidirectional E-glass woven fabrics aligned in the same direction, with a total weight of 216 grams. To create the adhesive mixture, 196 grams of epoxy resin were combined with 20 grams of the chosen hardener, triethylenetetramine, at a 1:10 weight ratio. This adhesive mixture was used for bonding the stacked plies together. During the hand lay-up process, the fabric plies were methodically positioned within the mold, and the adhesive mixture was applied between each ply. Pressure was subsequently applied to ensure effective bonding. Once the hand-layup procedure was completed, the composite laminates were subjected to curing. This was achieved by applying compression pressure using a dead weight on top of the mold. The curing process took place at room temperature for a duration of 48 hours. Subsequently, a post-cure process was conducted in an oven at a temperature of 80°C for a period of 2 hours [29]. To achieve the desired dimensions, the composite laminates were then cut to the required size using the water jet cutting technique.

## 2.3. The treatment and characterization of bonding surfaces

To ensure optimal bonding, a comprehensive two-step surface preparation procedure was applied to the substrates. Initially, a pre-grinding machine was employed to eliminate a significant portion of the inclined surface materials. This step effectively leveled the substrates, resulting in a more uniform surface. The primary purpose of this pre-grinding stage was to rectify any notable irregularities or inclinations present on the substrates.

Following the initial pre-grinding process, further refinement of the inclined surfaces was achieved through the use of sandpapers. A selection of four distinct sandpapers featuring varied grit sizes

(60, 240, and 800) was utilized to achieve a range of surface roughness profiles. During the sanding process, which extended over a duration of 10 minutes, the treatment was executed in a randomized manner. This approach ensured consistent and uniform treatment of the inclined surfaces, contributing to the overall effectiveness of the surface preparation. Following the completion of the sanding procedure, the substrates proceeded to a degreasing phase involving the use of acetone for a duration of 10 minutes. This degreasing method proved highly effective in the elimination of any potential contaminants or oils present on the surface. By doing so, it established a clean and receptive bonding surface for the subsequent application of adhesive. To further enhance the removal of any remaining debris or particles, a thorough rinsing process with distilled water was administered to the substrates. This meticulous approach ensured the attainment of impeccably clean and uncontaminated bonding surfaces [9]. Moreover, in order to facilitate comparison, certain specimens were subjected solely to the degreasing process using acetone, succeeded by a subsequent rinse with distilled water. The surface morphology of the fabricated laminated composite substrates was analyzed using scanning electron microscopy (SEM) with an acceleration voltage of 20 kV, performed using a VEGA TESCAN system from the Czech Republic. While SEM images offer valuable insights into surface topography, they do not provide a precise assessment of surface roughness. To accurately gauge the surface roughness, parameters such as the arithmetic average height (Ra), ten-point height (Rz), and root-mean-square roughness (Rq) were quantified using a profilometer (TR-200Plus). Five distinct points on each sample were evaluated, (as illustrated in Fig. 1). To enhance reliability and statistical significance, three repetitions were conducted for each experimental condition. This replication accounted for inherent variations, bolstering the robustness of the results. The average surface roughness parameters for each group were determined by calculating the mean of all 15 measurements.

To establish a robust bond and adhesion between the laminated composite substrates and the adhesive, a bonding surface characterized by high wettability and surface free energy is essential [30]. The static contact angles on the different samples were determined using a goniometer (VCA Optima by AST Products, Inc) under controlled conditions of 25°C temperature and 30% humidity. Equipped with a digital camera, the contact angle analyzer captured images of liquid droplets on the sample surfaces, and these images were subsequently analyzed using SE2500 software for precise contact angle measurements.

For calculating the surface energy through the Owens-Wendt method [31], two distinct liquids were employed for contact angle assessments: distilled water, representing the polar liquid, and diiodomethane, serving as the non-polar liquid. The properties of these liquids are outlined in Table 2. By employing the surface tension equilibrium equation of the solid-liquid-gas triple point (Eq. 2), the total surface free energy ($\sigma_s$) along with its components encompassing the polar ($\sigma_s^P$) and dispersive ($\sigma_s^D$) aspects were determined.

$$\gamma_{sl} = \sigma_s + \sigma_l - 2(\sqrt{\sigma_s^D \times \sigma_l^D} + \sqrt{\sigma_s^P \times \sigma_l^P}) \tag{1}$$

$$\sigma_s = \gamma_{sl} + \sigma_l \times \cos\theta \tag{2}$$

In the context of the equation, $\theta$ signifies the contact angle, $\gamma_{sl}$ represents the surface tension at the interface between the solid and liquid phases, and $\sigma_s$ and $\sigma_l$ denote the surface free energy of the substrate and the surface tension of the liquid, respectively. The superscripts D and P indicate the dispersive and polar components, respectively, while the total surface energy encompasses the sum of both these components.

For every individual sample, a droplet measuring 35 microliters was meticulously positioned onto the surface, maintaining a distance of 20 millimeters. Five distinct measurement points were identified for each droplet. This process was replicated three times for each group, leading to a cumulative count of 15 measurements per group. The arrangement of the measurement points is depicted in Figure 1.

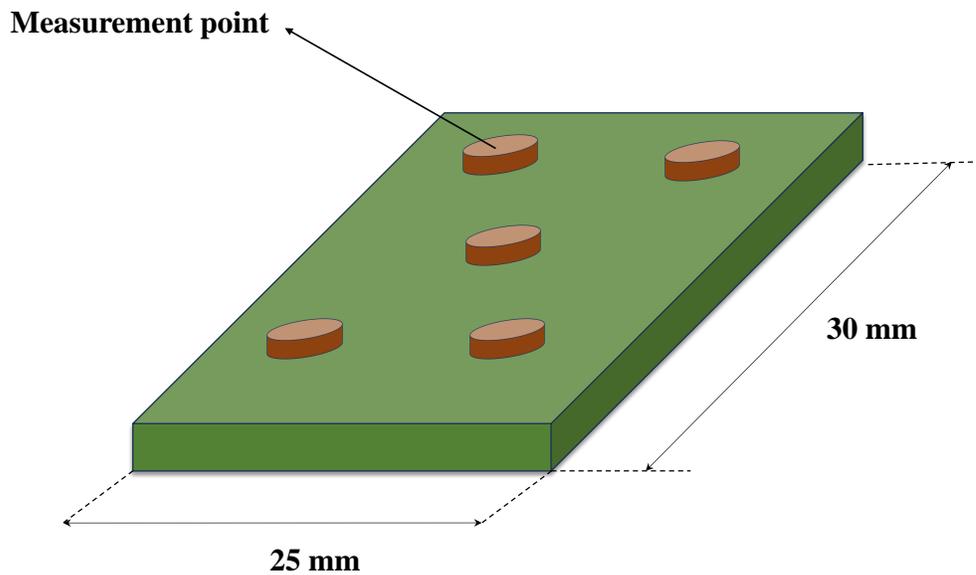

**Fig.1.** Measurement points distribution on the bonding surface.

Table 2. Surface tension components of liquids [31].

| Liquid | Polar component (mJ/m$^2$) | Dispersive Component (mJ/m$^2$) | Surface energy (mJ/m$^2$) |
|---|---|---|---|
| Distilled Water | 51.0 | 21.8 | 72.8 |
| Diiodomethane | 2.4 | 48.6 | 51.0 |

**2.4. Fabrication and testing of double-cantilever beam samples**

The DCB specimens were produced by connecting two composite substrates using the same resin that was used in the construction of the composite laminates. To initiate a pre-crack, a 15 µm thick aluminum foil was placed in the center of the adhesive thickness. For maintaining a precise adhesive layer thickness of 0.6 mm, two wires with a diameter of 0.6 mm were inserted between the substrates at the joint's ends. A manufacturing fixture was utilized to apply controlled pressure on the joints during the curing process. Afterward, any excess adhesive was carefully removed from the samples. The curing procedure was performed in an oven at a temperature of 60°C for a duration of 90 minutes [29]. MWCNTs were incorporated to reinforce the specimens. The process employed to fabricate the reinforced specimens closely resembled that used for creating the unreinforced specimens, with the exception that the nanofillers were introduced into the adhesive layer. In this research, MWCNTs were incorporated into the adhesive at three different weight percentages: 0.1, 0.3, and 0.5. The process of incorporating nanofillers involved initially mixing them into the epoxy resin using a mechanical stirrer at 500 rpm for 5 minutes. Subsequently, the homogeneous mixture was subjected to bath sonication at 50 kHz frequency and 180 watts power, maintained at a temperature of 40°C for 2 hours [24].

Both the reinforced and unreinforced DCB samples underwent testing following the guidelines outlined in ISO 25217 [32] to ascertain their mode-I fracture energy (G1C). The dimensions of the DCB specimens utilized for the testing procedure can be observed in Figure 2.

The quasi-static tensile tests were carried out using the SANTAM STM-150 universal testing machine with a displacement rate of 2.5 mm/min, maintaining displacement control. The experimental setup for the DCB specimens can be observed in Figure 3. Throughout the testing process, load-displacement curves were recorded for each specimen. The crack length during the tests was measured using a traveling optical microscope. The determination of G1C values, corresponding to the initiation of crack propagation, was conducted through the application of the simple beam theory.

A minimum of five specimens were subjected to testing in each experimental scenario to calculate the average G1C values. The fracture surfaces of the adhesive joints were carefully examined using a scanning electron microscope (SEM, VEGA/TESCAN) with an acceleration voltage of 20 kV. In preparation for SEM examination, the samples were mounted onto aluminum stubs and coated with a thin layer of Platinum. This coating process ensured optimal visualization of the samples under SEM analysis.

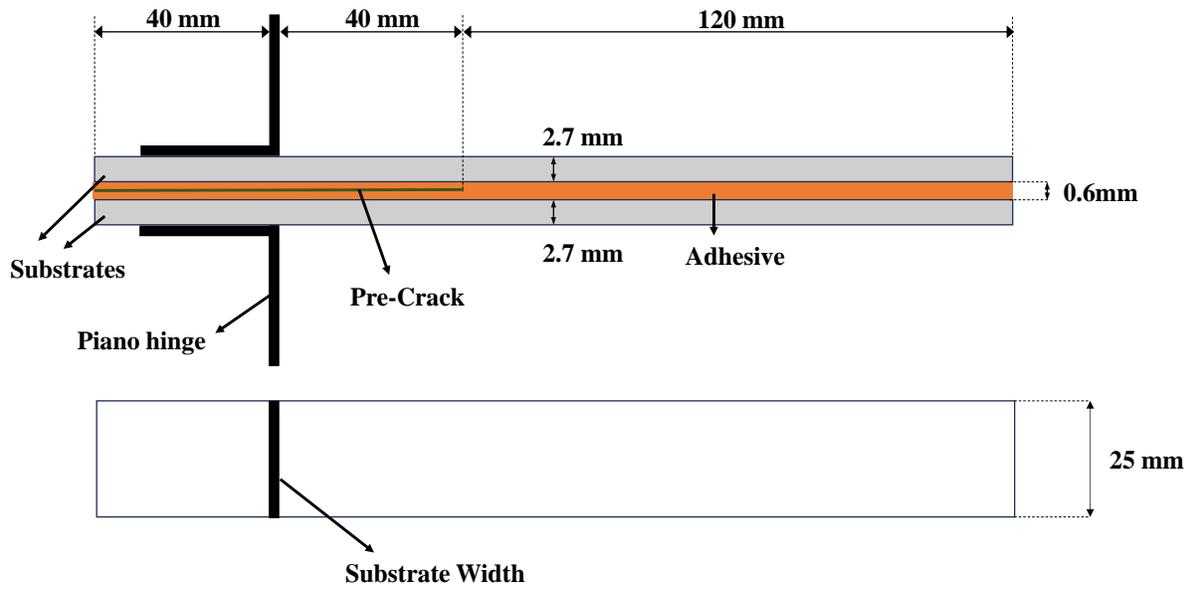

**Fig 2.** Schematic of the DCB specimens.

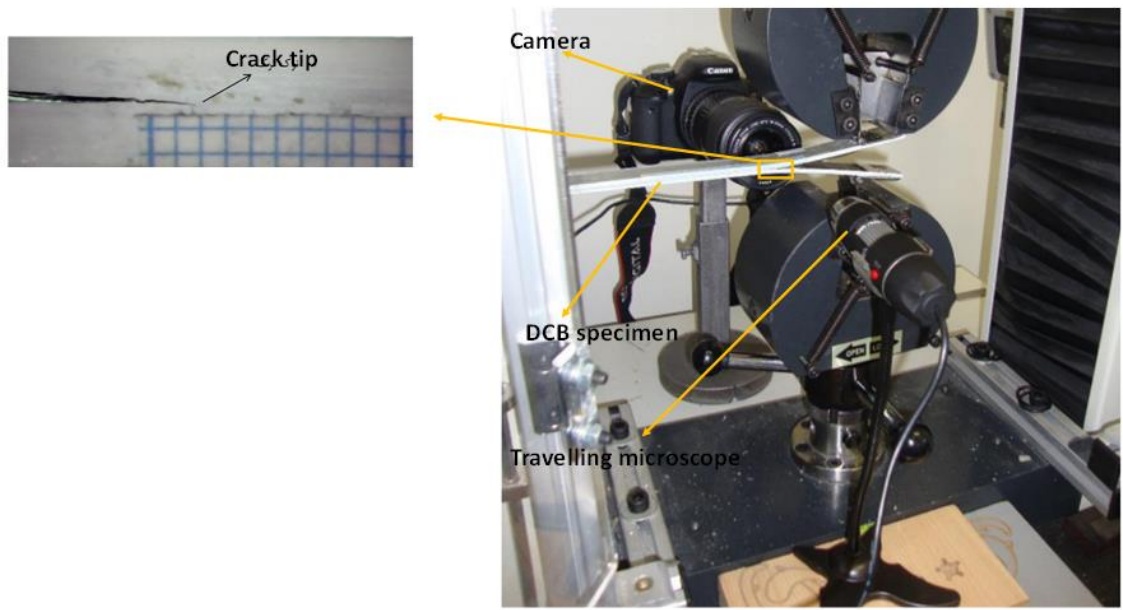

**Fig 3**. Test setup for the DCB specimens.

## 3. Results and Discussion

### 3.1. Load-displacement responses

The load-displacement behavior of the composite adhesive joints was evaluated in accordance with the established ISO 25217 [32] standard. Figure 4 illustrates the characteristic load-displacement curves of both the unreinforced composite adhesive joints and those enhanced with randomly distributed MWCNTs. These specimens were exposed to different surface treatment methods and incorporated various weight percentages of MWCNTs. Furthermore, Figure 5 presents a comparison of the average maximum load levels achieved for both the unreinforced specimens and those reinforced with varying weight percentages of MWCNTs. Upon close examination of both the unreinforced and reinforced specimens, several noteworthy observations can be made. Firstly, the load-displacement curves, as illustrated in Figure 5, display fluctuations as the load surpasses its peak value. This phenomenon is indicative of unstable crack propagation during failure, often referred to as stick-slip behavior [30].

Secondly, focusing on the unreinforced cases, the specimens treated with 240 grit size sandpaper exhibited the highest maximum load levels. Notably, these specimens demonstrated an 80% improvement in maximum load compared to the degreased counterparts. Conversely, the specimens treated with 60 grit size sandpaper experienced a significant 50% decrease in their maximum load capacity. These findings align with similar outcomes reported in references [9,33], where optimal load-bearing capacity was observed for specimens with a moderate level of surface roughness.

Thirdly, the trend persists in the reinforced cases, with the highest load values achieved by specimens treated with 240 grit size sandpaper, and the lowest load values by those prepared with 60 grit size. Moreover, the incorporation of MWCNTs with varying weight percentages contributed to enhanced maximum load levels in comparison to the unreinforced specimens. For example, the unreinforced specimens achieved a maximum load of 79.8 N when prepared with 240 grit size sandpaper. In contrast, the reinforced specimens with 0.1 and 0.3 wt % MWCNTs demonstrated maximum load levels of 101.5 N and 119.4 N, respectively, representing improvements of 27 % and 50 %. However, for specimens reinforced with 0.5 wt % MWCNTs, a modest 4 % improvement was observed.

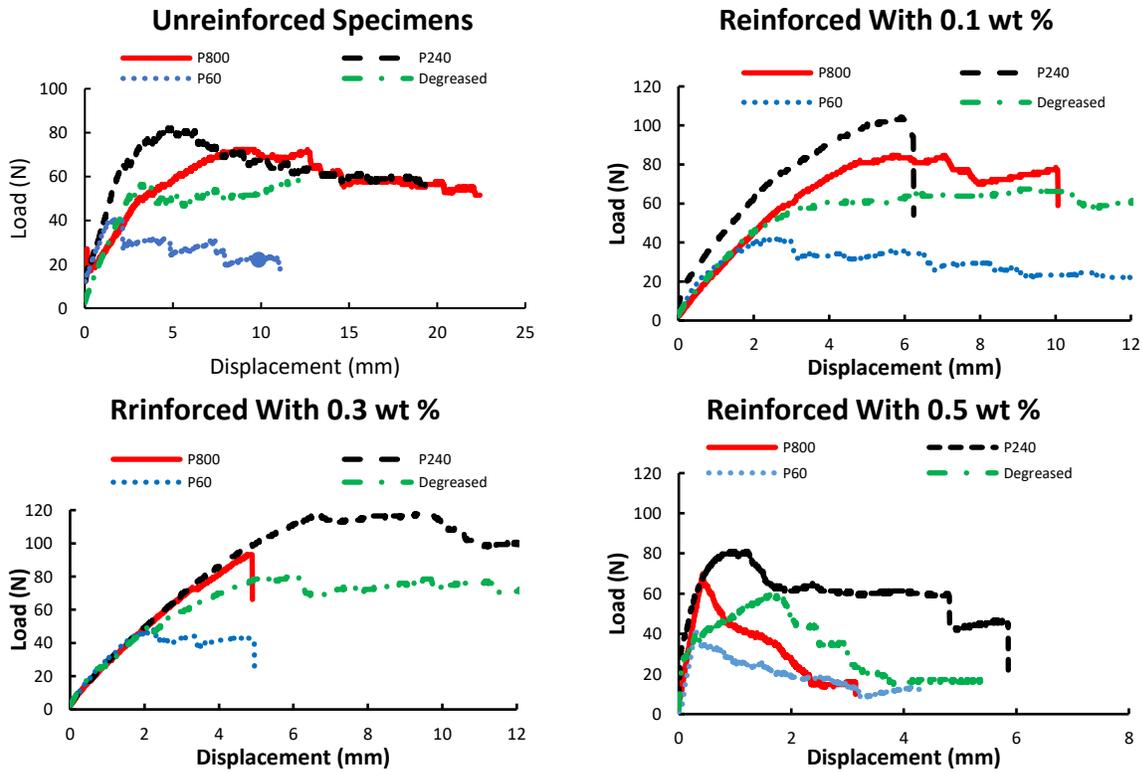

**Fig. 4.** Load–displacement responses of reinforced and unreinforced specimens.

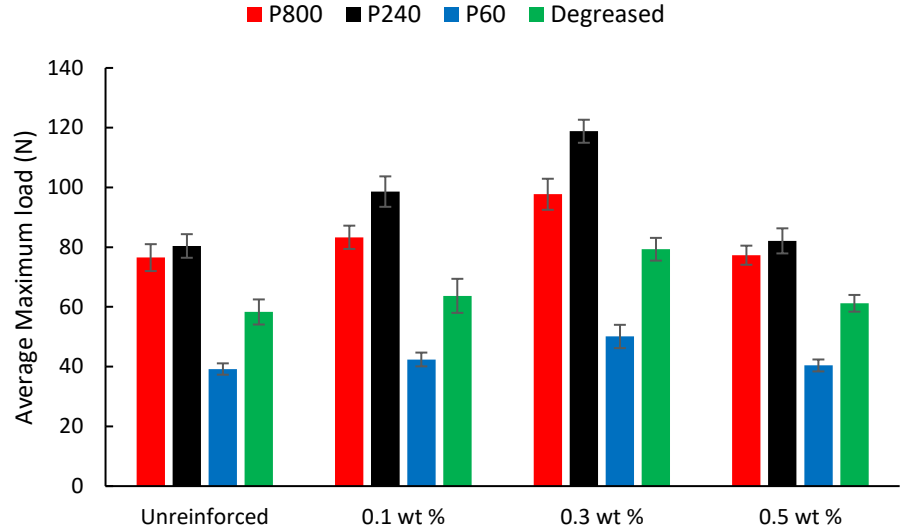

**Fig. 5.** Comparison between the average peak loads of reinforced and unreinforced specimens.

### 3.2. Fracture Energy

#### 3.2.1 Fracture energy at crack initiation

The fracture energy values at the onset of crack initiation were determined for both the unreinforced and reinforced composite adhesive joints according to ISO 25217. Based on standard ISO 25217, simple beam theory (SBT) can be employed for the determining mode-I fracture energy (G1C) when the crack growth follows a stick-slip pattern. Therefore, as in this study stick-slip pattern was observed as the dominant crack growth scheme, the simple beam theory was utilized for obtaining the fracture energy. According to the simple beam theory, the fracture energy (G1C) is obtained using Eq. (3).

$$G_{IC} = \frac{4\,P^2}{E B^2}\left(\frac{3a^2}{h^3}+\frac{1}{h}\right) \quad (3)$$

where P is the applied force, E is the elasticity modulus of substrates, b is the substrate width, h is the substrate thickness and a is the crack length.

Fracture energy values were calculated by identifying the load level corresponding to the initiation of crack growth, as defined by Eq. (3). Figure 6 depicts the mean fracture energy values (G1C) initiation for the composite adhesive joints. The composite adhesive joints were reinforced with randomly dispersed MWCNTs, each having different weight percentages and undergoing various surface treatments. This approach aimed to investigate the influence of these nanomaterials on the fracture behavior of the joints. Figure 6 clearly illustrates that the fracture energy values of the composite adhesive joints, which were reinforced with MWCNTs, exhibited a significant increase compared to the unreinforced specimens. Notably, the maximum initiation fracture energy for the unreinforced specimens, achieved through abrasion with 240 grit size, amounted to 1.14 N/mm. In contrast, the reinforced specimens, particularly at a 0.3 mixing ratio with 240 grit size, demonstrated a remarkable initiation fracture energy of 2.20 N/mm, signifying a noteworthy 92 % enhancement in initiation fracture energy when compared to the unreinforced specimens. For the reinforced specimens at 0.1 and 0.5 wt % with 240 grit size, an improvement of 71 % and 4 % was observed, respectively, compared to the unreinforced specimens prepared with the same grit size.

Moreover, both the unreinforced and reinforced specimens with various mixing ratios exhibited improved initiation fracture energy when subjected to sanding with 240 and 800 grit sizes, in contrast to the degreased specimens. However, the use of a 60 grit size resulted in decreased initiation fracture energy compared to the degreased specimens for both unreinforced and reinforced cases.

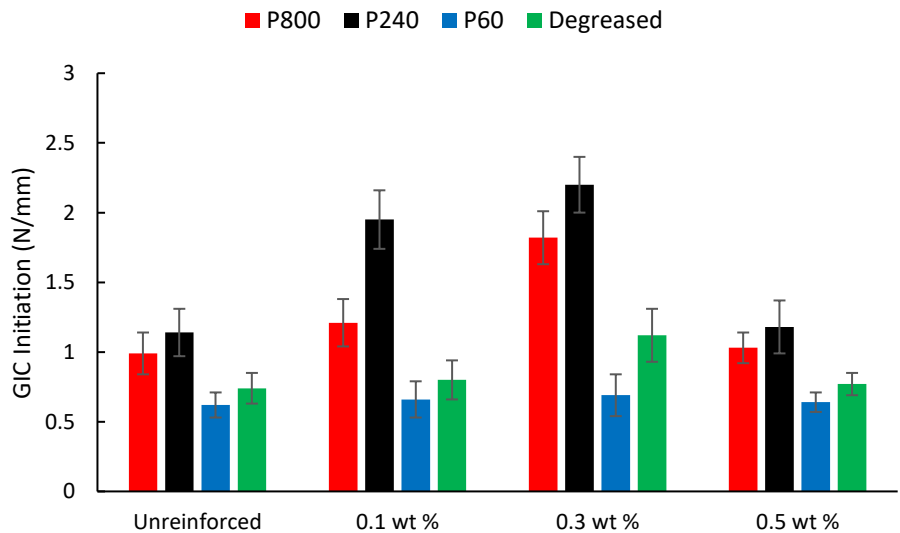

**Fig.6.** The initiation fracture energy of the unreinforced and the reinforced specimens.

### 3.2.2. Fracture energy and crack resistance response during crack growth

The fracture energy values during crack growth were determined for both the unreinforced and reinforced composite adhesive joints. by considering the crack length recorded during crack growth and the corresponding load level using Eq. (3). Figure 7 presents a comparison of the mean fracture energy values among different specimens as crack growth occurs.

As depicted in Figure 7, the propagation fracture energy values associated with crack propagation exhibited a similar trend to that observed for the maximum load and fracture energy at crack initiation. Reinforcing the composite laminated joints after undergoing sanding pretreatment with varying grit sizes led to substantial enhancements in the G1C (fracture energy at crack propagation) value, when compared to the unreinforced specimens. The findings demonstrated notable improvements of 57 %, 73 %, and 5 % for the reinforced specimens with 0.1, 0.3, and 0.5 wt % MWCNTs respectively, treated with 240 grit size, as compared to the unreinforced specimens subjected to the same grit size treatment. The results were in line with existing literature, where the addition of 0.3 wt % MWCNTs consistently yielded the highest fracture energy values [24,29].

Furthermore, among the unreinforced specimens, the highest G1C propagation was achieved for the specimens treated with 240 grit size sanding, resulting in an improvement of 80% compared to degreased specimens. Similarly, the specimens treated with 800 grit size sanding exhibited a 35% enhancement in $G_{IC}$ propagation compared to degreased specimens. However, the specimens treated with 60 grit size sanding showed a 10% decrease in G1C propagation. This trend was also observed for the reinforced specimens with different weight percentages.

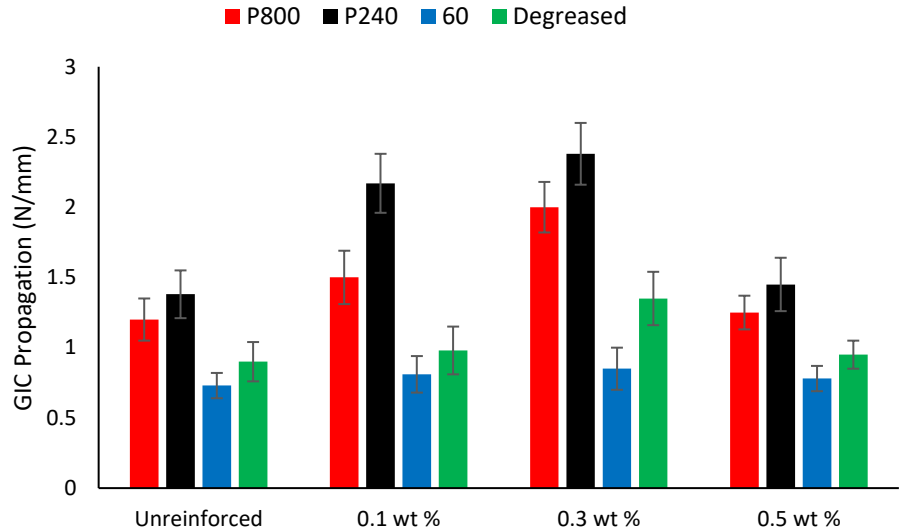

**Fig.7.** The initiation fracture energy of the unreinforced and the reinforced specimens.

## 3.3. Evaluation of the characteristics of the bonding surfaces

### 3.3.1 Bonding surface morphology and roughness

The surface morphology of the bonding areas in GFRP samples subjected to sanding and degreasing techniques was examined using SEM images. Figure 8 displays the SEM images of GFRP surfaces treated with various sandpaper grit sizes and the GFRP surfaces treated with degreasing. As the grit size of the sandpaper increases from 60 to the degreased method, the bonding surface displays shallower scratches and a smoother appearance at a macroscopic level. Additionally, within this specified range, the microscopic texture of the surface demonstrates a noticeable reduction in elevated peaks, resulting in the emergence of well-defined regular undulations on a micro scale. Table 3 presents the average surface roughness measurements across various bonding surfaces. Clearly, there is a significant reduction in surface roughness among the samples, ranging from 2.30 µm to 0.56 µm, as the grit size of the sandpaper increases.

In a system involving the bonding of materials, the mechanical interlocking effect is notably affected by specific crucial factors when the adhesive properly wets the substrates. These factors encompass the surface roughness, porosity, and unevenness of the bonding surface [9,34]. Surface porosity signifies the concentration of elevated and lowered points, while irregularity pertains to the surface's structural characteristics. Moreover, an excess of surface roughness can hinder the appropriate diffusion of the adhesive, thereby impacting the overall efficacy of the bonding process. As depicted in Figure 8, the degreased surfaces exhibited a comparatively smoother texture, with the glass fibers being completely enveloped by a layer of resin, as observed in Figure 8(a). The SEM images (Figure 8-c-e) illustrate that treatments with higher grit sizes produced bonding surfaces characterized by shallower peaks and valleys, fostering superior surface integration. This facilitated a more effective mechanical interlocking, ultimately resulting in an augmentation of the joint's load-bearing capacity and fracture resistance. However, as depicted in Figure 8(b), the act of abrading the bonding surfaces using 60-grit size sandpaper led to the removal of resin atop the fibers and the disruption of the fibers themselves, leading to the formation of pronounced grooves and valleys. These grooves are challenging to completely fill with adhesive, thereby creating areas of stress concentration at the interface [35]. This phenomenon consequently decreased the fracture energy and load-bearing capability of the joints.

**Table 3.** The bonding surface roughness parameters.

| Sandpaper grit sizes | $R_a$ (µm) | $R_q$ (µm) | $R_z$ (µm) |
|---|---|---|---|
| **Degreased** | 0.56±0.11 | 0.78±0.15 | 3.12± 0.26 |
| **P60** | 2.30±0.28 | 2.76±0.38 | 12.86±0.41 |
| **P240** | 1.55±0.17 | 1.94±0.25 | 10.13±0.50 |
| **P800** | 0.62±0.14 | 0.86±0.22 | 3.86±0.30 |

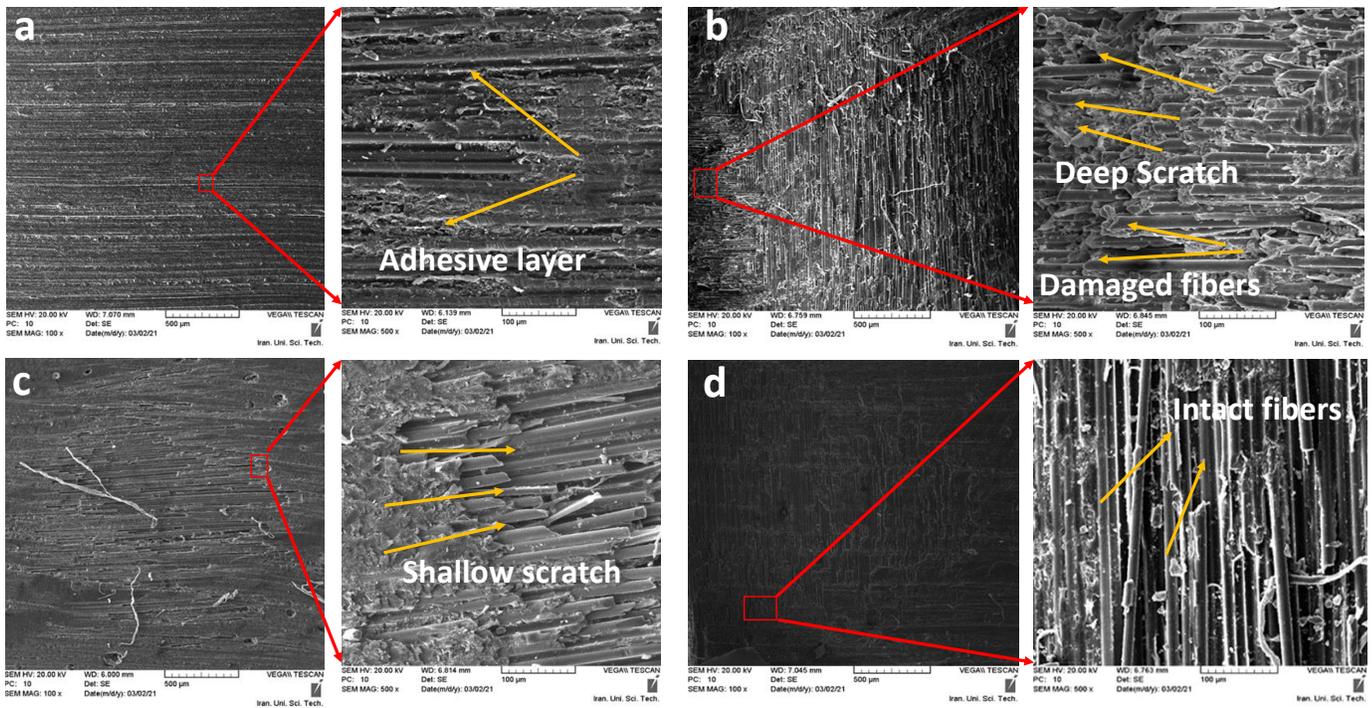

**Fig. 8.** SEM images of the surfaces treated with a) degreasing method and abraded with different sandpaper grit sizes of b)60, c)240, and d)800. Left images (100 x magnification) and Right images (500 x magnification)

### 3.3.2. Contact angle and surface free energy

Surface wettability can be characterized through measurements of surface energy and contact angles. The average contact angle and surface energy of the bonding surfaces for different specimens are illustrated in Figure 9 and 10, respectively. The collected data reveals a noteworthy pattern: the surface contact angles of the samples exhibit an initial decrease, followed by a gradual increase as the grit sizes increase. This trend in surface contact angles can be attributed to the influence of surface roughness. Interestingly, the surface energy exhibited a contrasting trend when compared to the surface contact angle. As depicted in Figure 9, the contact angle for the degreased specimens measured at 92.1 degrees, while the surface free energy was determined to be 37 mJ/m². Conversely, the contact angles for all surfaces treated with sandpaper were higher, and their corresponding surface free energies were comparatively lower when compared to those of the degreased surfaces (except 60 grit size). This observation of decreased contact angles and elevated surface free energies on the abraded bonding surfaces implies an enhanced level of wettability for these surfaces. Based on the adhesive adsorption theory, a higher surface free energy within specimens is associated with stronger bonding forces with external substances, thereby contributing to improved bonding effects. Consequently, according to this theory, it is expected that the load bearing capacity and fracture energy of the samples would initially increase and subsequently decrease, moving from the specimens treated with a grit size of 240 (which exhibits the lowest contact angle of 76.5 and 54.8 degrees for the distilled water and diiodomethane respectively, and the highest surface free energy of 44 N/mm²) to the degreased specimen.

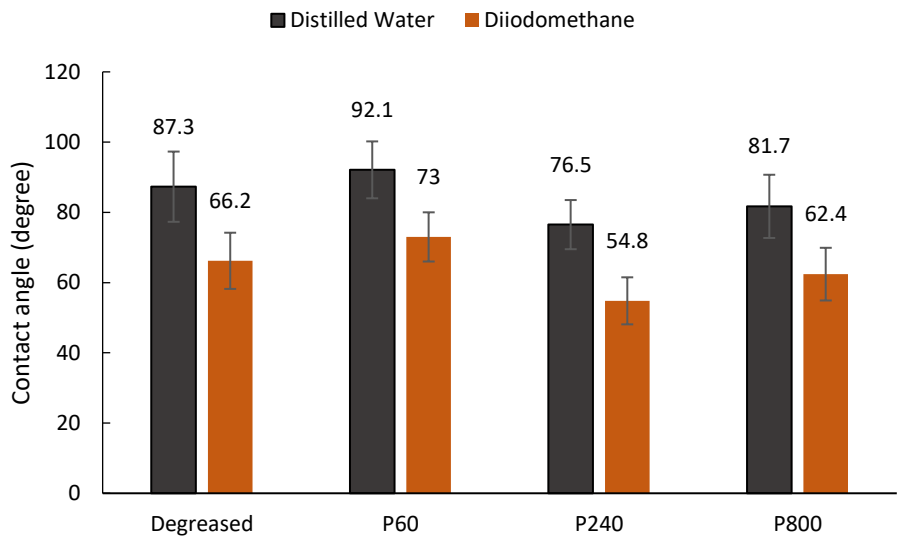

**Fig. 9.** The contact angles of various liquid on different specimens.

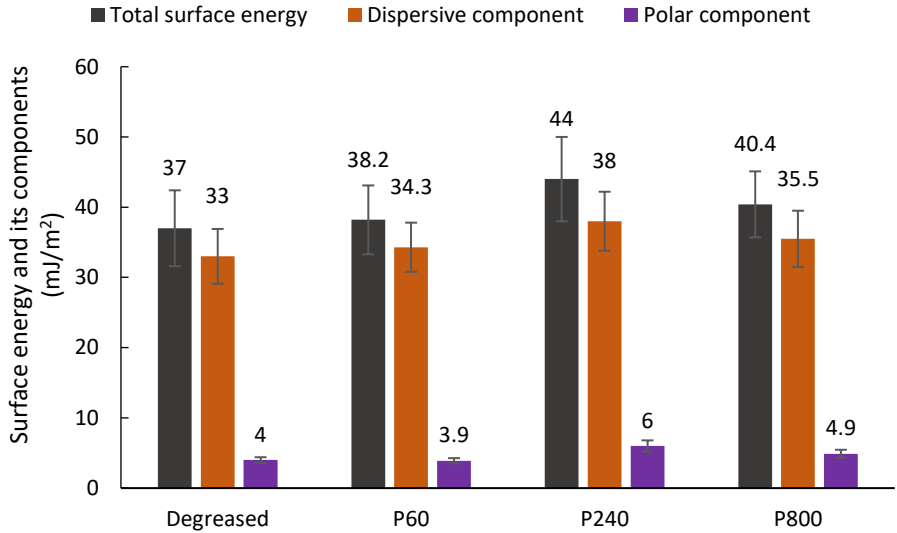

**Fig. 10.** Total surface free energy and its components

### 3.4. Toughening Mechanism

SEM images were employed to investigate the toughening mechanisms induced by MWCNTs. Fig. 11 displays typical SEM images showcasing the fracture surfaces of the composite adhesive joints that were reinforced with varying weight percentages of MWCNTs. As observed in Figure 11(a), the fracture surface of the unreinforced composite adhesive joint appears relatively smooth, indicating a brittle fracture with low fracture toughness [29]. However, with the addition of nanofillers, it's evident that the roughness of the fracture surfaces increased. Upon dispersing rigid nanofillers within the epoxy resin, the presence of these particles led to crack deflection when encountered by growing cracks. Consequently, cracks were compelled to deviate from their initial propagation path due to tilting and/or twisting effects caused by the rigid nanoparticles [4]. Furthermore, the emergence of fracture surfaces with increased roughness demands higher energy during the process of crack propagation. In the SEM images depicting the specimens containing 0.5 wt% MWCNTs (Figure 11-d), it is evident that the nanofillers were not effectively dispersed, resulting in visible agglomerated nanoparticles. These agglomerations were primarily responsible for diminishing the fracture resistance, particularly at elevated weight percentages of nanofillers. Agglomerated nanofillers have the potential to create local stress concentrations within the adhesive layer, which, in turn, can diminish the overall fracture resistance of the adhesive joint [36]. This observation aligns with the outcomes depicted in maximum load and fracture energy graphs (Figure 5-6). When the specimens were enhanced with 0.5% MWCNTs, only marginal improvements of 4% in maximum load and 4% in initiation fracture energy were achieved in comparison to the unreinforced specimens at same grit size. Conversely, reinforcement at 0.1 and 0.3 wt% led to substantial enhancements of 27 % and 50 % in maximum load, and 71 % and 92 % in initiation fracture energy, respectively, when compared to the unreinforced specimens. Aside from crack deflection, the pull-out and crack bridging mechanisms represent additional toughening mechanisms apparent on the fracture surfaces of the reinforced adhesive joints containing MWCNTs (as illustrated in Figure 12). These mechanisms provide insights into the elevated $G_{IC}$ values computed for the adhesive joints reinforced with MWCNTs in comparison to the unreinforced adhesive. The nanofillers acted as bridges between the two surfaces of the crack as it propagated. Under continued loading, the nanofillers were drawn out of the surfaces, contributing to an increase in fracture energy [24,37].

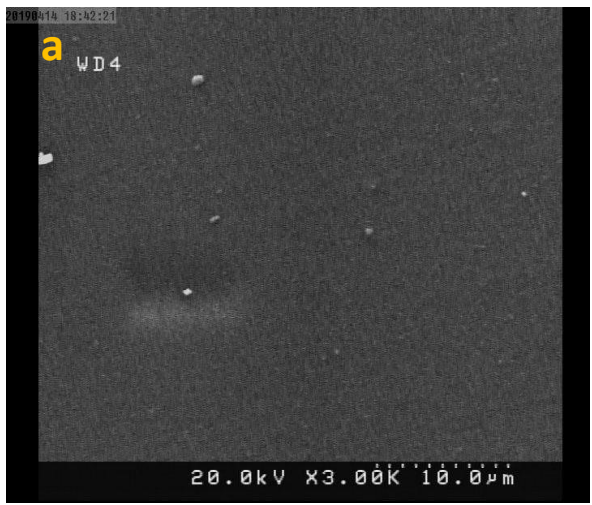
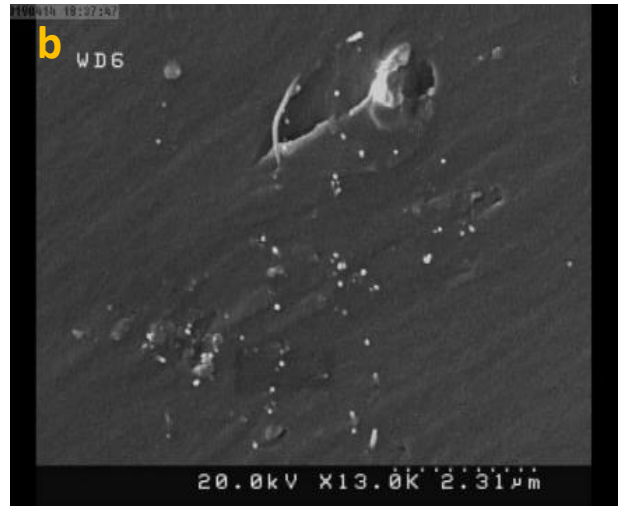

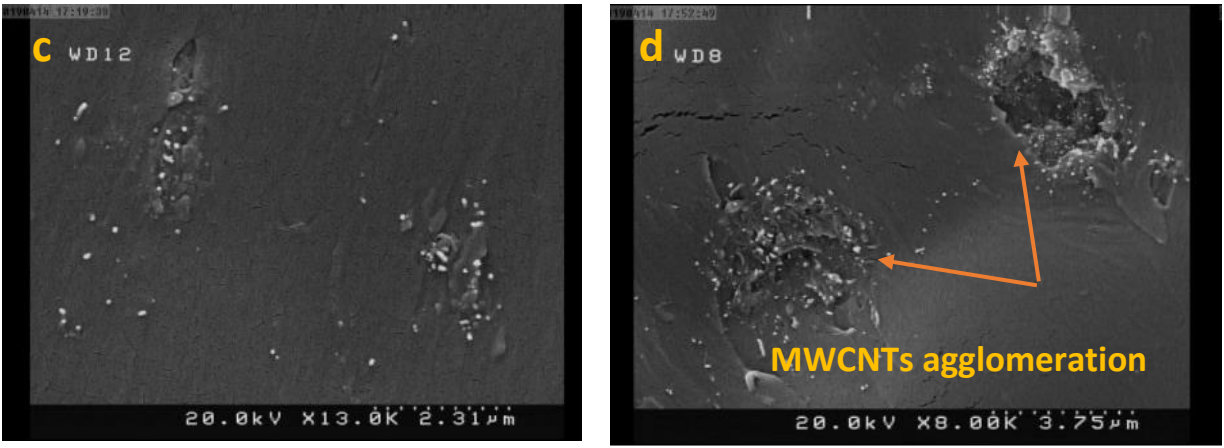

**Fig.11.** FE-SEM images captured from the fracture surfaces of a) unreinforced specimens and reinforced with b) 0.1 wt % c) 0.3 wt % and d) 5 wt % MWCNTs.

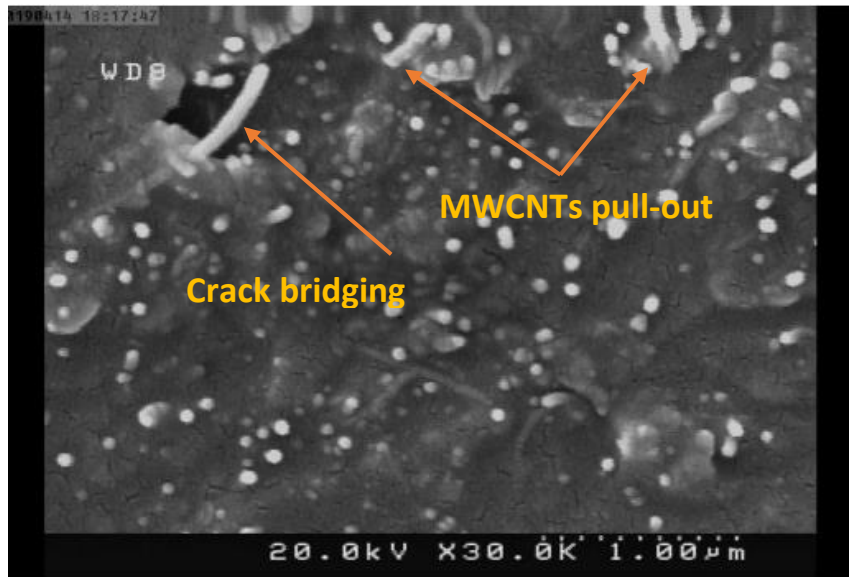

**Fig.12.** FE-SEM images captured from the fracture surfaces of reinforced with 0.3 wt % MWCNTs

### 3.5. Fracture surface

The examination and analysis of fracture surfaces in composite adhesive joints were carried out using the ISO 10365 standard [38]. This standard was utilized to ascertain and comprehensively study the distinct patterns of failure exhibited by these fracture surfaces. Three distinct failure patterns were identified upon examining the fracture surfaces of the specimens. In the first observed pattern, damage was concentrated within the adhesive layer itself, with cracks propagating exclusively within this layer. This particular mode was termed cohesive failure (CF). In a separate failure pattern, the point of failure was localized at the boundary between the adhesive and the composite substrates, leading to what was termed adhesive failure (AF). The third failure pattern involved failure transpiring at the interface connecting the fabric and matrix within the initial layer of the composite substrates adjacent to the adhesive layer. This specific mode was designated as cohesive substrate failure (CSF). Figure13 displays the fracture surfaces of both unreinforced composite adhesive joints and those reinforced with MWCNTs. The distinct failure patterns observed in each specimen are outlined in Table 4.

As depicted in Figure 13, in the case of the unreinforced adhesive joints where the bonding surfaces underwent solely a degreasing process, the prevailing failure mode was identified as adhesive failure. This suggests a deficiency in adhesion between the adhesive and the composite substrates. This occurrence can be rationalized by the remarkably smooth nature of the substrate's surface, which hindered the escape of gas molecules. Consequently, these gas molecules were confined within the adhesive layer during the bonding process, in contrast to rougher surfaces where air within grooves or gaps could be effortlessly released [18]. Abrading the bonding surfaces resulted in a notable shift in the failure mode, transitioning towards cohesive failure (P240). This shift entailed the propagation of cracks occurring within the adhesive layer itself, as opposed to the interface between the substrate and adhesive. This change in behavior can be ascribed to the gradual increase in the surface roughness of the bonding surfaces. With an increase in surface roughness, more pronounced and deeper scratches emerged, consequently fostering a more consistent and uniform dispersion of the adhesive layer [9]. However, employing a coarser sandpaper (with a grit size of 60) for abrading the bonding surfaces led to a reversion of the failure mode back to interfacial, as illustrated in Figure 13(b). This shift pointed to a diminishment in the quality of bonding. The underlying cause for this observation can be attributed to the impairment inflicted upon the surface fibers of the specimens that underwent sanding with the grit size of 60. The abrasive grains effectively incised these fibers, rendering them more susceptible to detachment from the substrate during the DCB test. This outcome aligns with previous research where the degradation of surface fibers was identified as a factor that compromised adhesive joints, ultimately culminating in a failure mode characterized by the tearing of fibers [21,35,39].

As outlined in Table 4, the introduction of various weight percentages of MWCNTs as reinforcements for the adhesive joints resulted in a progressive alteration of the failure pattern. This evolution was observed as a transition from Cohesive Failure (CF) and Cohesive Substrate Failure (CSF) patterns. This trend signifies the successful reinforcement of the adhesive joints. Furthermore, it indicates that the adhesive layer's fracture resistance surpassed that of both the adhesive-substrate interface and the fabric-matrix interface within the composite structure. To provide an example, consider the specimens reinforced with 0.3 wt % of the reinforcement and subjected to sanding using a 240 grit size. In this scenario, the observed failure mode was Cohesive Substrate Failure (CSF). This can be attributed to a synergistic effect arising from two factors. First, the application of a moderate grit size of 240 for sanding improved the interfacial bond between the adhesive layer and the substrate. Second, the reinforcement of the adhesive layer with 0.3 wt % contributed to an enhancement in the fracture resistance of the adhesive layer itself. These combined influences resulted in the observed CSF failure mode.

The shifts in the failure mode of the joints align closely with the findings derived from the analysis of bonding surface characteristics and the outcomes of the adhesive joint fracture tests. Among the unreinforced specimens that underwent abrasion with 60-grit size sandpaper, the recorded total surface free energy was the lowest, accompanied by the highest contact angle. Coinciding with these characteristics, the resulting adhesive joints displayed the lowest values for maximum load and fracture energy during the conducted tests. In terms of failure mode, the outcome was characterized by an adhesive failure pattern. In contrast, the specimens reinforced with 0.3 wt % and subjected to abrasion using 240-grit size sandpaper exhibited the highest total surface free energy. Concurrently, they displayed the lowest contact angle. These attributes coincided with the adhesive joints produced from these specimens, which achieved the highest values for maximum load and fracture energy during testing. Furthermore, the failure mode observed in these cases was cohesive substrate failure.

**Table.4.** The dominant failure patterns observed for the composite adhesive joints.

| Specimens | Degreased | P60 | P240 | P800 |
|---|---|---|---|---|
| Unreinforced | AF | AF | CF | CF+ AF |
| 0.1 wt % | CF | CF+AF | CSF | CF |
| 0.3 wt % | CF | CF | CSF | CF+CSF |
| 0.5 wt % | CF | CF+AF | CF | CF |

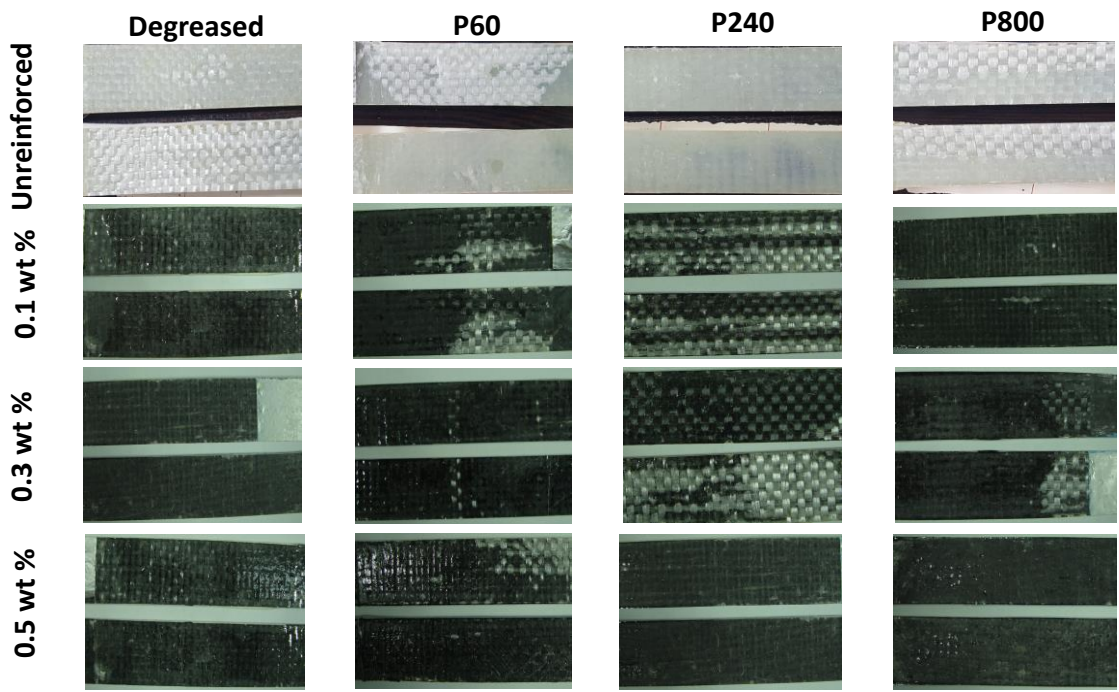

**Fig.13.** Typical fracture surfaces of adhesive joints

4. **Conclusion:**

In this study, the impact of mechanical abrasion techniques and the incorporation of Multi-Walled Carbon Nanotubes (MWCNTs) on bonding surface attributes and fracture behavior of adhesive joints involving glass fiber-reinforced polymer (GFRP) composites was systematically examined. This investigation was conducted through the evaluation of Double Cantilever Beam (DCB) specimens in accordance with the ISO 25217 standard.

Considering of the unreinforced specimens, the research revealed a notable observation: the application of sandpaper with an optimal grit size can induce substantial alterations in the bonding surface attributes and lead to enhanced fracture behavior of adhesive joints. In this context, various sandpaper grit sizes, spanning from 60 to 800, were utilized to prepare the bonding surfaces of the joints. Specifically, when the sandpaper grit size was decreased from 800, resulting in heightened surface roughness, the contact angle diminished. Simultaneously, the total surface free energy experienced an increase, eventually reaching an optimal value upon employing the 240-sandpaper grit size. These findings were consistent with the outcomes of the fracture tests conducted on the joints, where various sandpaper grit sizes were employed to prepare the bonding surfaces. Specifically, the adhesive joint that was prepared using 240-grit size sandpaper (representing the optimal case) demonstrated a substantial enhancement. It experienced a remarkable 38 % increase in the maximum load threshold and a significant 55 % improvement in mode-I fracture energy, in comparison to the joint where only degreasing of bonding surfaces was performed. Nevertheless, when the sandpaper grit size was further decreased to 60, the consequences were quite the opposite. The joint's maximum load capacity experienced a substantial decline of 32 %, while the mode-I fracture energy saw a reduction of 19 %. These figures were in stark contrast to the joint that underwent only degreasing of its bonding surfaces.

The introduction of various weight percentages of MWCNTs as reinforcements for the adhesive joints resulted in an improvement in the fracture behavior composite laminated joints. The investigation revealed that specimens reinforced with 0.3 wt % and subjected to abrasion using 240-grit size sandpaper demonstrated the most elevated values for maximum load and fracture energy. This enhancement equated to an impressive increase of 50 % and 92 % respectively, when compared to the unreinforced specimens that underwent degreasing. In this context, the detected failure mode was identified as Cohesive Substrate Failure (CSF). This phenomenon can be explained by the interaction of two contributing factors. Firstly, the utilization of a moderate 240-

grit sandpaper size enhanced the bonding between the adhesive layer and the substrate interface. Secondly, the incorporation of 0.3 wt % reinforcement into the adhesive layer augmented the fracture resistance of the adhesive material itself. The collaborative impact of these two factors culminated in the observed CSF failure mode.

**References:**

[1] Pramanik, A., et al. (2017). "Joining of carbon fibre reinforced polymer (CFRP) composites and aluminium alloys–A review." Composites Part A: Applied Science and Manufacturing 101: 1-29.

[2] Hamdi, M., M.N. Saleh, and J.A. Poulis, Improving the adhesion strength of polymers: effect of surface treatments. Journal of Adhesion Science and Technology, 2020. 34(17): p. 1853-1870.

[3] Nazari, R. and H. Khoramishad (2022). "A novel combined anodic-adhesive bonding technique for joining glass to metal for micro device applications." International Journal of Adhesion and Adhesives 117: 103175.

[4] Ahmadi-Moghadam, B. and F. Taheri, Fracture and toughening mechanisms of GNP-based nanocomposites in modes I and II fracture. Engineering Fracture Mechanics, 2014. 131: p. 329-339.

[5] Guo, L., et al. (2021). "Effects of surface treatment and adhesive thickness on the shear strength of precision bonded joints." Polymer Testing 94: 107063.

[6] Budhe, S., et al. (2017). "An updated review of adhesively bonded joints in composite materials." International Journal of Adhesion and Adhesives 72: 30-42.

[7] Jeevi, G., S. K. Nayak and M. Abdul Kader (2019). "Review on adhesive joints and their application in hybrid composite structures." Journal of Adhesion Science and Technology 33(14): 1497-1520.

[8] Liu, J., et al. (2023). "Review of the surface treatment process for the adhesive matrix of composite materials." International Journal of Adhesion and Adhesives 126: 103446.

[9] Yang, G., et al., The influence of surface treatment on the tensile properties of carbon fiber-reinforced epoxy composites-bonded joints. Composites Part B: Engineering, 2019. 160: p. 446-456.

[10] Tao, R., M. Alfano, and G. Lubineau, Laser-based surface patterning of composite plates for improved secondary adhesive bonding. Composites Part A: Applied Science and Manufacturing, 2018. 109: p. 84-94.

[11]    Wang, Dawei, et al. "Increasing strength and fracture toughness of carbon fibre-reinforced plastic adhesively bonded joints by combining peel-ply and oxygen plasma treatments." Applied Surface Science 612 (2023): 155768.

[12]    Buchmann, Christopher, et al. "Analysis of the removal of peel ply from CFRP surfaces." Composites Part B: Engineering 89 (2016): 352-361.

[13]    Lee, S.-O., K.Y. Rhee, and S.-J. Park, Influence of chemical surface treatment of basalt fibers on interlaminar shear strength and fracture toughness of epoxy-based composites.

[14]    Sun, Chengcheng, et al. "Effect of atmospheric pressure plasma treatment on adhesive bonding of carbon fiber reinforced polymer." Polymers 11.1 (2019): 139.

[15]    Zaldivar, R., et al., The effect of atmospheric plasma treatment on the chemistry, morphology and resultant bonding behavior of a pan-based carbon fiber-reinforced epoxy composite. Journal of composite materials, 2010. 44(2): p. 137-156.

[16]    Lim JD, Yeow SYS, Rhee MWD, Kam CL, Chee CW. Surface roughness effect on copper–alumina adhesion. Microelectron Reliab 2013;53(9–11):1548–52

[17]    Baldan A. Adhesion phenomena in bonded joints. Int J Adhesion Adhes2012;38(4):95–116.

[18]    Zhan X, Li Y, Gao C, Wang H, Yang Y. Effect of infrared laser surface treatment onthe microstructure and properties of adhesively CFRP bonded joints. Optic LaserTechnol 2018;106:398–409.

[19]    Gude MR, Prolongo SG, Ureña A. Adhesive bonding of carbon fibre/epoxy laminates: correlation between surface and mechanical properties. Surf Coating Technol2012;207(9):602–7.

[20]    Kumar, S., et al., Effect of atmospheric pressure plasma treatment for repair of polymer matrix composite for aerospace applications. Journal of Composite Materials, 2016. 50(11): p. 1497-1507.

[21]    Sun, F., et al., Influence of surface micropatterns on the mode I fracture toughness of adhesively bonded joints. International Journal of Adhesion and Adhesives, 2020. 103: p. 102718.

[22]	Martínez-Landeros, V., et al., Studies on the influence of surface treatment type, in the effectiveness of structural adhesive bonding, for carbon fiber reinforced composites. Journal of Manufacturing Processes, 2019. 39: p. 160-166.

[23]	Khoramishad H and Zarifpour D. Fracture response of adhesive joints reinforced with aligned multi-walled carbon nanotubes using an external electric field. Theor Appl Fract Mech 2018; 98: 220–229.

[24]	Radshad, H., H. Khoramishad and R. Nazari (2022). "The synergistic effect of hybridizing and aligning graphene oxide nanoplatelets and multi-walled carbon nanotubes on mode-I fracture behavior of nanocomposite adhesive joints." Proceedings of the Institution of Mechanical Engineers, Part L: Journal of Materials: Design and Applications 236(9): 1764-1776.

[25]	Ayatollahi MR, Moghimi Monfared R and Barbaz Isfahani R. Experimental investigation on tribological properties of carbon fabric composites: effects of carbon nanotubes and nano-silica. Proc IMechE Part L: J Materials: Design and Applications 2019; 233: 874–884.

[26]	Kumar Srivastava A and Kumar Pathak V. Elastic properties of graphene-reinforced aluminum nanocomposite: effects of temperature, stacked, and perforated graphene. Proc IMechE Part L: J Materials: Design and Applications 2020; 234: 1218–1227.

[27]	Gude M, Prolongo S, Gómez-del Río T, et al. Mode-I adhesive fracture energy of carbon fibre composite joints with nanoreinforced epoxy adhesives. Int J Adhes Adhes 2011; 31: 695–703.

[28]	Khoramishad H and Khakzad M. Toughening epoxy adhesives with multi-walled carbon nanotubes. J Adhes 2018; 94: 15–29.

[29] Gholami R, Khoramishad H and da Silva LFJCS. Glass fiber-reinforced polymer nanocomposite adhesive joints reinforced with aligned carbon nanofillers. Compos Struct 2020; 253: 112814.

[30]	Tzetzis, D. and P. Hogg, the influence of surface morphology on the interfacial adhesion and fracture behavior of vacuum infused carbon fiber reinforced polymeric repairs. Polymer Composites, 2008. 29(1): p. 92-108.

[31] Rudawska, A. and E. Jacniacka, Analysis for determining surface free energy uncertainty by the Owen–Wendt method. International journal of adhesion and adhesives, 2009. 29(4): p. 451-457.

[32] Normalización, O.I.d., Adhesives: Determination of the Mode 1 Adhesive Fracture Energy of Structural Adhesive Joints Using Double Cantilever Beam and Tapered Double Cantilever Beam Specimens. 2009: ISO.

[33] Sun C, Min J, Lin J, Wan H, Yang S, Wang S. The effect of laser ablation treatmenton the chemistry, morphology and bonding strength of CFRP joints. Int J AdhesionAdhes 2018;84:325–34.

[34] Baldan A. Adhesion phenomena in bonded joints. Int J Adhesion Adhesive2012;38(4):95–116.

[35] Kim KS, Yoo JS, Yi YM, Kim CG. Failure mode and strength of uni-directionalcomposite single lap bonded joints with different bonding methods. Compos Struct2006;72(4):477–85

[36] Khoramishad H and Hosseini Vafa S. Effect of aligning graphene oxide nanoplatelets using direct current electric.

[37] Chandrasekaran S, Sato N, Tölle F, et al. Fracture toughness and failure mechanism of graphene based epoxy compo sites. Compos Sci Technol 2014; 97: 90–99.

[38] ISO E. 10365. Adhesives–designation of the main failure patterns. Europe: BSI Standards, 1992.

[39] Roohollah Nazari, Hadi Khoramishad & Rashid Hakimi (2024) The effect of abrading treatment method on the bonding surface characteristics and fracture behavior of glass fiber-epoxy composite adhesive joints, Journal of Adhesion Science and Technology, DOI: 10.1080/01694243.2024.2302259